\documentclass{aastex63}
\usepackage{subfiles} 
\usepackage{comment}
\usepackage{xcolor}

\chardef\us=`\_

\received{January 6, 2021}
\revised{April 11, 2021}
\accepted{\today}
\submitjournal{ApJ}

\shorttitle{Flare Forecasting Algorithms Based on High-Gradient PILs}
\shortauthors{Cicogna et al.}
\graphicspath{{./}{figures/}}

\begin{document}

\title{Flare Forecasting Algorithms Based on High-Gradient Polarity Inversion Lines in Active Regions}

\correspondingauthor{Francesco Berrilli}
\email{francesco.berrilli@roma2.infn.it}

\author{Domenico Cicogna}
\affiliation{Department of Physics, University of Rome Tor Vergata, Via della Ricerca Scientifica, 1, I-00133, Rome, Italy}
\affiliation{NEXT Ingegneria dei Sistemi S.p.A., Rome, Italy}

\author[0000-0002-2276-3733]{Francesco Berrilli}
\affiliation{Department of Physics, University of Rome Tor Vergata, Via della Ricerca Scientifica, 1, I-00133, Rome, Italy}

\author[0000-0003-2755-5295]{Daniele Calchetti}
\affiliation{Department of Physics, University of Rome Tor Vergata, Via della Ricerca Scientifica, 1, I-00133, Rome, Italy}

\author[0000-0003-2500-5054]{Dario Del Moro}
\affiliation{Department of Physics, University of Rome Tor Vergata, Via della Ricerca Scientifica, 1, I-00133, Rome, Italy}

\author[0000-0001-7369-8516]{Luca Giovannelli}
\affiliation{Department of Physics, University of Rome Tor Vergata, Via della Ricerca Scientifica, 1, I-00133, Rome, Italy}

\author[0000-0002-4776-0256]{Federico Benvenuto}
\affiliation{Dipartimento di Matematica, Università di Genova, via Dodecaneso 35, I-16146 Genova, Italy}

\author[0000-0003-2105-8554]{Cristina Campi}
\affiliation{Dipartimento di Matematica, Università di Genova, via Dodecaneso 35, I-16146 Genova, Italy}

\author[0000-0001-7047-1148]{Sabrina Guastavino}
\affiliation{Dipartimento di Matematica, Università di Genova, via Dodecaneso 35, I-16146 Genova, Italy}

\author[0000-0003-1700-991X]{Michele Piana}
\affiliation{Dipartimento di Matematica, Università di Genova, via Dodecaneso 35, I-16146 Genova, Italy}



\begin{abstract}
Solar flares emanate from solar active regions hosting complex and strong bipolar magnetic fluxes. Estimating the probability of an active region to flare and defining reliable precursors of intense flares is an extremely challenging task in the space weather field.
In this work, we focus on two metrics as flare precursors, the unsigned flux R, tested on MDI/SOHO data and one of the most used parameters for flare forecasting applications, and a novel topological parameter D representing the complexity of a solar active region.\\ 
More in detail, we propose an algorithm for the computation of the R value which exploits the higher spatial resolution of HMI maps. This algorithm leads to a differently computed R value, whose functionality is tested on a set of cycle 24th solar flares.
Furthermore, we introduce a topological parameter based on the automatic recognition of magnetic polarity-inversion lines in identified active regions, and able to evaluate its  magnetic topological complexity. We use both a heuristic approach and a supervised machine learning method to validate the effectiveness of these two descriptors to predict the occurrence of X- or M- class flares in a given solar active region during the following 24 hours period.
Our feature ranking analysis shows that both parameters play a significant role in prediction performances. Moreover, the analysis demonstrates that the new topological parameter D is the only one, among 173 overall predictors, which is always present for all test subsets and is systematically ranked within the top-ten positions in all tests concerning the computation of the weighs with  which each  predictor impacts the flare forecasting.

\end{abstract}


\section{Introduction}
     \label{S-Introduction} 
     Solar flares represent the most energetic explosive events in our solar system.
     They consist of sudden and powerful coronal events which are triggered by plasma instabilities and energized by reconnections of the magnetic field present in coronal loops \citep[e.g.][]{PF2002}, which in turn are controlled by their foot-points advection by plasma flows in photosphere \citep[e.g.][]{Viticchie2006,Reale2014,Caroli2015,Kuridze2019}.
     Their most impressive manifestation is the sudden release of large amounts of energy, mainly emitted in the high frequency regions (e.g., above extreme ultraviolet, EUV) of the solar electromagnetic spectrum (see \citealt{Fletcher2011,2019LRSP} and references cited therein).
     These bursts of energy, in addition to producing a consistent increase in the high energy radiative flux, are often accompanied by intense fluxes of solar energetic particles, mainly electrons and protons, and sometimes by Coronal Mass Ejections (CMEs) \citep[e.g.][]{Chen2017}.
     Energetic particles, CMEs, 'frozen-in' solar magnetic field, and high-energy radiation cross the interplanetary space and interact with the magnetosphere and upper atmosphere of the planets or with artificial satellites that are in their path, generating the physical processes associated with circumterrestrial and planetary space weather \citep[e.g.][]{Dorman2005, Plainaki2016}.
     
     Indeed, solar energetic particles and high-energy radiation pose severe threats to satellites operating in higher orbits than Low-Earth Orbits and, more importantly, to astronauts operating outside the Earth's protective magnetosphere especially if involved in extravehicular activity \citep[e.g.][]{narici2018solar,Walsh2019}.
     Moreover, the increase of today’s society complexity and dependence on space technology represents a relevant risk factor as many of the technologies used, in space or ground-based assets, are potentially vulnerable to solar flares \citep[e.g.][]{Cannon2013, difino2014, berrilli2014relativistic}. 
     Operators of satellites, air lines, railways or power grids are continuously confronted by a number of challenges, ranging from mitigating the effects of solar flares and CMEs on their assets to the growing need of carefully predicting these events.
     For this reason, many national or international institutions have developed forecasting services within large space weather projects \citep[e.g.][]{ESA2010, NOAA2012, IPS2019A, IPS2019B, plainaki2020} or research groups have worked on targeted flare or multi-flare forecasting tools \citep[e.g.][]{Barnes2007, Schrijver2007, GR2007, falconer2011, Kors2015, Romano2015, Florios2018, McCloskey2018, Lim2019, SWERTO2019, Falco2019, Giovannelli2019, Yi2020, Nishizuka2020, Kors2020, Cinto2020}.
     
     Usually, flare forecasting algorithms described in the literature use an approach based on the physics of the processes necessary to trigger solar flares. In practice, those algorithms analyze Line-of-Sight (LoS) or vector magnetograms, or other maps of physical quantities such as LoS velocity or spectral intensity, of the recognized active regions to estimate single value quantities (i.e., descriptors) on the basis of several image analysis techniques \citep[e.g.][]{Campi2019}.
     Forecasts based on the extracted descriptors usually provide the probability of occurrence of a flare of a given class within 24 hours and require a robust statistical analysis on a large number of events (or non-events), flaring and non-flaring active regions, and for a long time period. 
     For this reason, different predictive statistical techniques have been applied, also including machine learning (ML) techniques \citep[e.g.][]{Barnes2016, Benvenuto2018, Piana2019, Campi2019, Barnes2019A, Barnes2019B, Barnes2020, benvenuto2020machine}.
     
    In this study we aim at assessing the flare forecasting performances of two descriptors estimated by the analysis of the LoS magnetograms associated to a set of active regions (ARs). These ARs are both flaring and non-flaring within 24 hours and are acquired by the Helioseismic Magnetic Imager on-board the Solar Dynamics Observatory (SDO/HMI) during solar cycle 24 (SC24).
     
     The first predictor is a suitable generalization of the morphological metric (R), originally proposed by \cite{Schrijver2007} for SOHO/MDI LoS magnetograms acquired during solar cycle 23 by the Michelson Doppler Imager on-board the Solar and Heliospheric Observatory (SOHO/MDI). 
          The second feature is a new topology-based metric (D) that estimates the magnetic complexity of the active region by counting the number of polarity inversion lines (PILs) above a suitable magnetic field threshold. 
     This metric, as we will discuss in detail in the following sections, represents an infrequent use of PILs as it focuses on the analysis of the magnetic topology, rather than the mere morphology, of potentially flaring regions. In fact, as discussed in \cite[e.g.][]{Antiochos1998}, the complexity of the photospheric magnetic field, defining the right topology for magnetic breakout, is an essential element for strong activity such as large flares.
     
     In the application of this analysis, the paper fulfills two main goals. The first one is to recalibrate the \cite{Schrijver2007} algorithm for the calculation of R, originally calibrated on LoS magnetograms with the resolution of SOHO/MDI, on LoS magnetograms at the full resolution of SDO/HMI. 
     This predictor, which is scaled to have the same range as R and therefore to be easily compared to the original Schrijver's feature, will be named R$^*$ in the following. 
     The second goal is to compare the performances of the two predictors, namely R$^*$ and D, by flaring and non-flaring scenarios in order to quantify their predictive capability and reliability.
     
     The content of the paper is as follows. 
     Section 2 describes the SDO/HMI dataset of flaring and non-flaring regions and the selection procedure. 
     Section 3 overviews the recalibration process of R Schrijver’s feature, explaining in detail the numerical procedure to compute the R value for SDO/HMI full resolution magnetograms and the calibration of original log R and the new log R$^*$. 
     Section 4 introduces the new topological descriptor D, shortly explains the conceptual background of this new feature, and explains in detail the numerical procedure to compute it. 
     Section 5 contains the results of the forecasting analysis using  R$^*$ and D, while our conclusions are offered in Section 6.

\section{The SDO/HMI dataset of flaring and non-flaring regions}
     \label{S-dataset}
The first task to perform for testing and calibrating R$^{*}$ and D was to create the magnetogram dataset of non-flaring and flaring regions during solar cycle 24.
A catalog of all the solar flares detected by the Geostationary Operational Environmental Satellites (GOES) is available from the Space Weather Prediction Center at the National Oceanic and Atmospheric Administration (SWPC/NOAA)\footnote{https://www.ngdc.noaa.gov/stp/satellite/goes/index.html}.
The GOES flare catalog provides, among the other information, the flare class, the flare start time and the NOAA active region number. Flares are detected in X-ray flux in the $0.1-0.8$ nm channel of
GOES ($F$) and their class is determined by the peak in such curve ($F_{peak}$).
We selected from the NOAA catalog the list of all M and X flares, from June 2010 to June 2018.
The magnetogram dataset was selected from the HMI Active Region Patches (HARPs) available from the Joint Science Operations Center (JSOC)\footnote{http://jsoc.stanford.edu/jsocwiki/HARPDataSeries} that provides LOS magnetograms (MHARPs) from the SDO/HMI instrument \citep{HMI} (specifically, we have used the hmi.sharp$\_$cea$\_$720s data stream). HARP pipeline identifies and tracks ARs in the solar photosphere generating a time series of magnetogram patches for each AR. 

%
The data set has been created by adopting the same criteria reported in \cite{Schrijver2007}. More in detail, we selected LOS magnetograms within 45 degrees from disk center for each day from June 2010 to June 2018. Limiting the dataset to near disk center reduce the influence of projection effects on LoS magnetograms. We then compared the magnetogram dataset with the GOES flare list to label magnetograms as flaring or non-flaring.

%
On the one hand, the flaring set contains only regions hosting events with peak flux greater than 10$^ {-5}$ Wm$^{-2}$ (M- or greater class flares).
We considered only ARs hosting a single flare, since in case of ARs flaring multiple times over a period of days, R remains high as magnetic flux continues to emerge \citep[see e.g.][]{Piersanti2017}.
This set is composed of 100 magnetograms of 100 different HARPs hosting an M- or X- class flare in the next 24 hours. 
On the other hand, the non-flaring set is composed by a collection of HARPs that did not produce any M- or X- class flare during their lifetime.
This set can contain more than one magnetogram per HARP, separated by at least 24 hours. This control group includes 745 total magnetograms.
This number of magnetograms, belonging to non-flaring regions, keeps the ratio between flaring and non-flaring regions equal to that used to calibrate R in \cite{Schrijver2007}. 
It is worth noting that the total statistics of the analyzed regions in this work is however different from that of \cite{Schrijver2007}, due to the different level of activity between cycle 23 and cycle 24.
The properties of the dataset are resumed in Table \ref{tab:dataset}.
\begin{table}[h]
    \centering
    \begin{tabular}{cc}
    \hline
    \textbf{Flaring HARPs}     &   \textbf{Non-Flaring HARPs}\\
    \hline
    100 magnetograms     &  745 magnetograms\\
    24 hours before peak time & Neither M- nor X- class flares\\
    No multiple flares  & \\
    One magnetogram per AR  &   24 hours separated magnetograms per AR\\
    45$^\circ$ from the center  &   45$^\circ$ from the center\\
    \hline
    \end{tabular}
    \caption{Description of the dataset used to characterize the R$^*$ parameter defined on the HMI/SDO magnetograms and the new topological parameter D. The ratio between the flaring AR magnetograms and the total number of entries is the same used by \cite{Schrijver2007}.}
    \label{tab:dataset}
\end{table}

In the following, we use this dataset for the computation of R$^*$ and D and to assess their forecasting capability.
Moreover, we also compare the two new descriptors with the properties developed by the FLARECAST Consortium (\url{http://flarecast.eu}) using Near-Realtime Space Weather HMI Archive Patch (SHARP) data \citep{Bobra2014}.

\section{SDO/HMI R$^{*}$ value algorithm}
Among the physics-based descriptors, i.e, among the descriptors that estimate the probability of the occurrence of a flare on the basis of physical properties derived from maps of the selected active region, the R value \citep{Schrijver2007} is of great significance.
The computation of R is based on the hypothesis that the emergence of electrical currents embedded in the magnetic flux is a key ingredient in the triggering of flares \citep[e.g.][]{Wheatland2000, schrijver2005}.
The presence of these currents is strongly linked to the presence of polarity-inversion lines (PILs) which are regions in which there is an intense 
vertical-component of the photospheric magnetic field ($B_n$) of opposite polarity.
Magnetograms from ground-based instruments \citep[e.g.][]{Cavallini2002, Cavallini2006, Scharmer2008, Steward2011, Puschmann2012, Tadesse2013, DelMoro2017JP3D, Forte2018, Forte2020, Giovannelli2020} or on-board satellites \citep[e.g.][]{Scherrer1995, Scherrer2012, Peter2012, Berrilli2015JATIS, Suematsu2017} can be used to detect PILs.
The R value is defined in \citep{Schrijver2007} and is a measure of the unsigned photospheric magnetic flux close to selected PILs, where an appropriate weighting map is applied for the SOHO/MDI instrument.\\
\begin{figure}[tbp]
\centering
\includegraphics[width=0.48\textwidth]{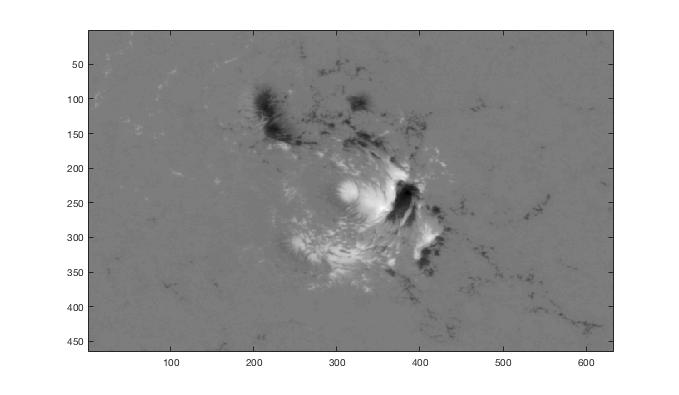}
\includegraphics[width=0.48\textwidth]{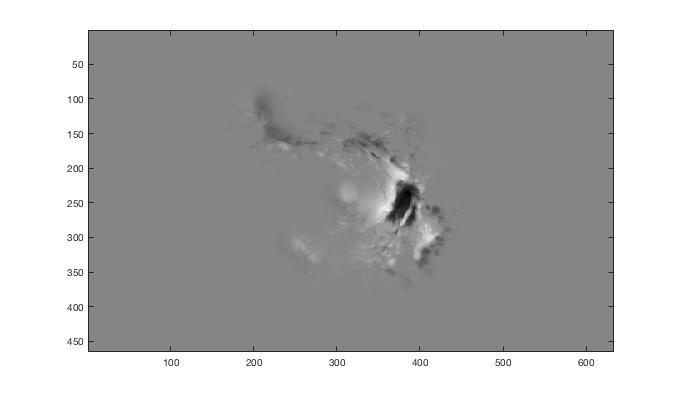}
\caption{Left: SDO HMI magnetogram of AR 12673 (NOAA number) on 5 September 2017, with an X9.3 flare within 24 hours. Right: HMI magnetogram multiplied for the weighting map. The summation over all pixels gives R$^*$.
All the axis are expressed in pixels.
}
\label{fig:Ralgorithm}
\end{figure}
The Michelson Doppler Imager \citep[MDI:][]{Scherrer1995} instrument was active from 1996 to April 2011 onboard the Solar and Heliospheric Observatory (SOHO) satellite. As a legacy of the MDI instrument, since 2010 SDO/HMI \cite{Scherrer2012} has been providing full vector solar magnetograms, with higher spatial and temporal resolution. 
Indeed, the cadence of the tracked Active Regions Patches for HMI is 12 minutes and  the images have a spatial resolution of 1 \textit{arcsec}, while MDI had a 96 minutes cadence and a spatial resolution equal to 4 \textit{arcsec}.
The pixel scale for HMI and MDI are 0.5 \textit{arcsec}/pixel and 2 \textit{arcsec}/pixel, respectively. The spectral line used for the computation of the vector magnetic field maps are different. MDI  used the Ni 676.8 nm line, while HMI uses the Fe I 617.3 nm line \citep{Norton2006}.
The line-formation heights are substantially similar, the models suggesting 125 km for the Fe I 617.3 nm line and 100 km for the Ni 676.8 nm line \citep{Fleck2011}.
Clearly, the differences in the experimental setup between the two instruments affect the magnetograms produced.
Their use for the calculation of the descriptors therefore requires an accurate calibration.
The comparison of the LoS magnetograms acquired during the overlapping period of the two missions revealed that the LoS magnetic signal inferred by MDI is larger than that derived from HMI by a factor of 1.40 \citep{liu2012}.
The same authors provided a recipe to convert HMI images to MDI-like images, adapting temporal cadence, spatial resolution and calibrating for the magnetic field strength.
In the present study we adapt the original \citep{Schrijver2007} algorithm, for the R value calculation, so that it can be applied to the full spatial resolution HMI magnetograms, and we benchmark its performance.
\begin{figure}[htbp]
\centering
\includegraphics[width=0.40\textwidth]{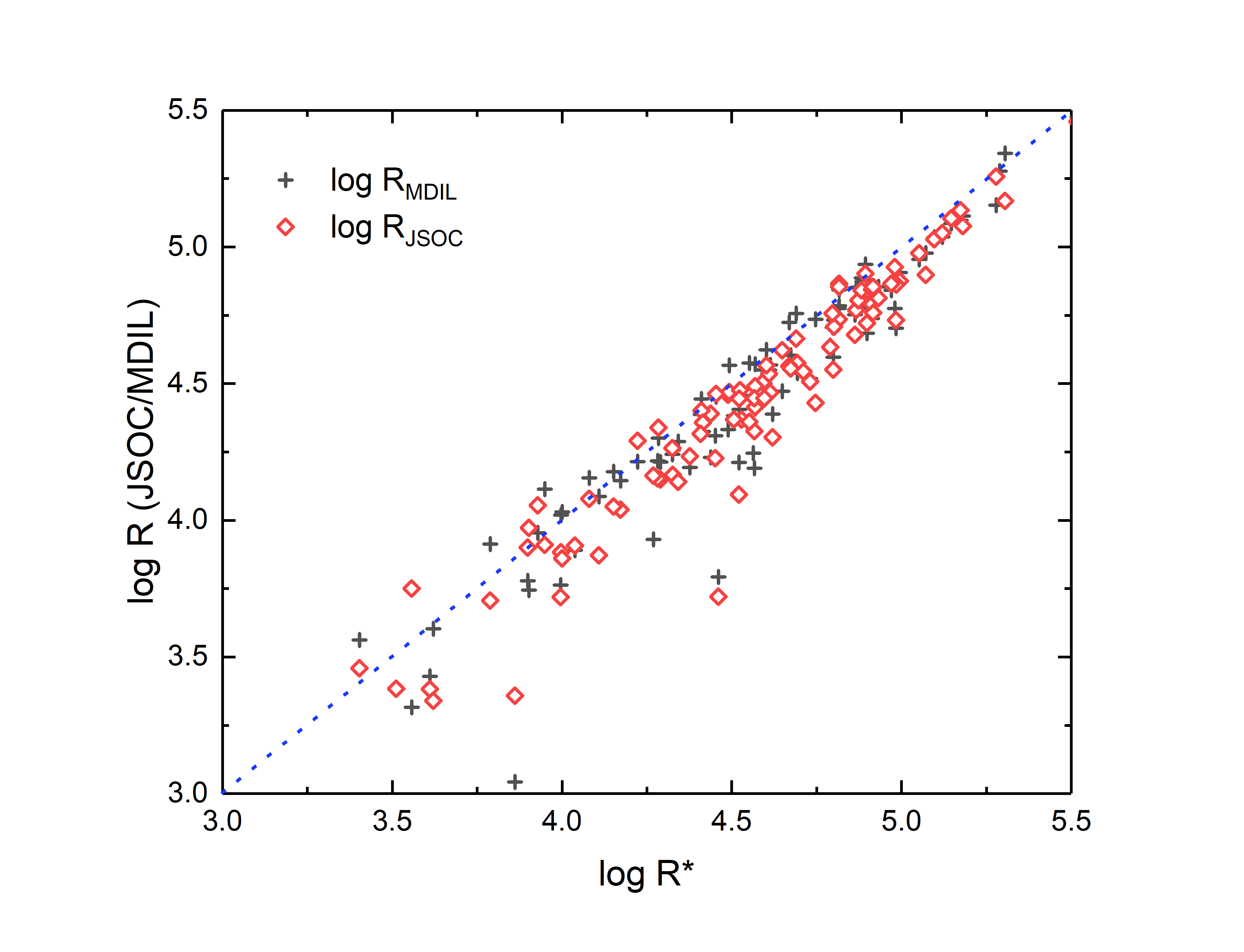}
\includegraphics[width=0.50\textwidth]{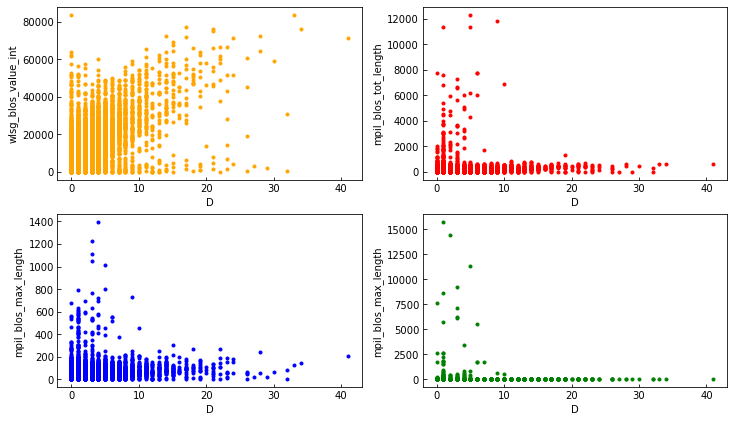}
\caption{Left panel: scatter diagram of the correlation between log R$_{\textrm{JSOC}}$, as obtained from the JSOC magnetograms header (red diamonds) and log R$_{\textrm{MDIL}}$, as obtained from reprocessing SDO/HMI magnetograms in MDI-like magnetograms, according to the recipe presented in \cite{liu2012}, and applying the \cite{Schrijver2007} algorithm (black crosses).
The points actually shown in this plot are 99, since an outlier with log R$_{MDIL}=0$ and log R$^{*}=1.8$ was excluded. Right panel: scatter diagrams of the correlation between $D$ and the features ``wlsg$\_$blos/value$\_$int'' (top left), ``mpil$\_$blos/tot$\_$length'' (top right), ``mpil$\_$blos/max$\_$length'' (bottom left), and ``mpil$\_$blos/tot$\_$usflux'' (bottom right). The definition of these descriptors can be found in Table \ref{tab:descriptors_definition}.
}
\label{fig:R_conf}
\end{figure}
We name the new descriptor R$^{*}$. The procedure and the tests to compute and benchmark R$^{*}$ are as follows.
First of all, we appropriately rescale the dilation kernels necessary to identify the high-gradient polarity-separation lines, maintaining the correct spatial scales. 
The bitmaps of HMI magnetograms, both for positive and negative magnetic flux densities, are dilated with kernels of $9 \times 9$ pixels (see Figure \ref{fig:Ralgorithm}).
To create the HMI bitmaps we keep the same threshold in flux densities (i.e., 150 G) originally used by \cite{Schrijver2007}, despite the difference in magnetic sensitivity of MDI and HMI instruments discussed above. 
The reason for this is that we want to consider only high gradient regions.

The two dilated bitmaps are then overlapped to find the pixels where their product is non-zero, showing the high-gradient polarity-separation lines.
Then, the bitmap containing the polarity-separation lines is convolved with a Gaussian function with a FWHM of $D_{sep} = 15$ Mm to calculate a weighing map.
The value of 15 Mm is the average maximum distance between the occurring flare in EUV images of the corona and the nearest point of any PILs in the region according to \cite{Schrijver2007}. 
This weighting map is then multiplied by the absolute value of the flux in the magnetograms.

Finally, in order to estimate the new value R$^{*}$, we take into account the difference in pixel area of MDI and HMI.
Indeed, the pixel areas are $\simeq 2.2$ Mm$^2$/pixel for MDI and $\simeq 0.14$ Mm$^2$/pixel for HMI, respectively. 
The R$^{*}$ value is eventually obtained from the weighted unsigned flux summed over all pixels.

There are various reasons why we think it is of interest to generalize the morphological metric R. First of all we can potentially take advantage of full resolution of the SDO/HMI magnetograms, but especially we can study the behavior of this parameter when the algorithm is tuned on instruments that have different spatial and spectral characteristics. This fact widens the applicability of the algorithm and is useful to more easily intercalibrate metrics derived by magnetograms produced by different instruments. Another important aspect is the replication of the statistical analysis, according to the procedure specified in \cite{Schrijver2007}, and the complementary supervised machine-learning analysis, in order to verify the behavior of the new metric on a second solar cycle. Applying the modified descriptor to a new solar cycle means extending the reliability analysis and confirming its forecasting capabilities.

A first validation of the R$^{*}$ descriptor has been performed by comparing the log R$^{*}$ values with the values of log R recorded in the Joint Science Operations Center (JSOC) data header, log R$_{JSOC}$, and with the estimate scaled to the resolution of SOHO/MDI magnetograms according to the recipe presented in \cite{liu2012}, log R$_{MDIL}$.
In Figure \ref{fig:R_conf}, left panel, we show the scatter plot of the values of log R$_{JSOC}$ and log R$_{MDIL}$ compared to log R$^{*}$ for the magnetograms of 100 different HARPs hosting an M or X class flare in the next 24 hours described in section \ref{S-dataset}.
Combining the log R values from these three different methods, we find a strong correlation between log R$_{JSOC}$ and log R$_{MDIL}$ and the estimations from our algorithm, log R$^{*}$.

\section{D value algorithm}
It has been known since a long time that the probability of occurrence of a flare is higher in magnetically complex active regions, e.g., Mount Wilson type $\gamma$ spot groups, compared to magnetically simple ones \citep[e.g.][]{Giovanelli1939, Howard1964, Antiochos1998}. For this reason, for example,  prediction algorithms based on a morphological measure of roughness, as the fractal or multifractal dimension, of the active regions have been proposed. However, these geometric properties have proved more useful in studying the scale properties of solar magnetism than in characterizing active regions in terms of eruptive ability or flare forecasting \citep[e.g.][]{Georgoulis2012}. The algorithms that measure morphological quantities related to the absolute magnetic vertical-field component {$B_n$} (approximated by the LoS field component near disk center) have turned out to be much more efficient in terms of predicting flares. Among these, in addition to the Schrijver's R value \citep{Schrijver2007}, that we have extensively discussed in the previous section, we find the WLSS and WLSG parameters of Falconer \citep{Falconer2009} or the procedures based on gradient-weighted inversion-line length \citep{MH2010}. As reported in \cite{Georgoulis2013} all of these parameters weigh heavily on PILs, although with different specifics.\\
The proposed D value, although based on the identification of PILs, is not based on morphological measures such as the partial or total measurement of the length of the PILs or a $B_n$ weighted estimate in the active region. Indeed, D is not a morphological descriptor, e.g., based on $B_n$ shape or distribution. On the contrary, it is a topological descriptor, therefore linked to quantitative measurement of magnetic features (the total number of PILs) in an active region. The D value, although very simple to calculate as we will discuss below, has similarities with the connectivity matrix, which is sensitive to flux partitioning, and which is based on a flux-tessellation scheme used for topological analyses \citep{Georgoulis2013} or with the connection and persistence diagram analysis used to evaluate the information about the spatial scales of the topological features in active regions \citep{Deshmukh2020}. 

As other complexity metrics, the topological complexity increases with the number of loops \citep[e.g.][]{Schwarz2013}, i.e., of PILs. If we also consider that the amount of magnetic flux near a PIL gives us an estimate of the reservoir of magnetic energy potentially available for the flare, it is natural to combine the two quantities and introduce a new parameter based on the number of PILs present in an AR.
With this motivation, we define this new metric to estimate the probability of triggering such an energy release.
Our assumption is that a complex and fragmented magnetic region could be more effective in accumulating free energy and triggering the flare process.

\begin{figure}[htbp]
\centering
\includegraphics[width=0.9\textwidth]{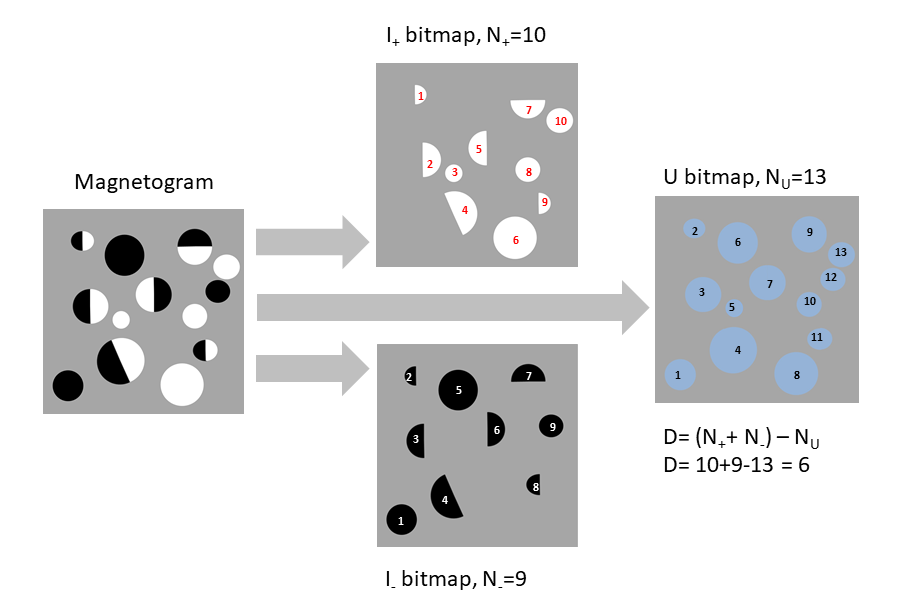}
\caption{The scheme for calculating the D value consists of few simple steps. First, a segmentation and labeling technique applied to the magnetogram identifies and counts the number of compact flux density structures in the three bitmaps, $N_{+}$, $N_{-}$, and $N_{U}$, respectively. Successively, the difference $D=(N_{+} + N_{-}) - N_{U}$ counts the number of PIL fragments in the magnetogram.}
\label{Dscheme}
\end{figure}

Therefore, the proposed topological descriptor D counts the number of separate PIL fragments present in the active regions, i.e. the number of different polarity-inversion lines in the $B_{LoS}$ magnetogram of the selected active region. 
Our implementation to calculate D exploits a fast thresholding and labelling procedure. 
Starting from the $B_{LoS}$ magnetogram $M(x,y)$ of the selected AR, indicated as Magnetogram in Fig.\Ref{Dscheme}, we apply an absolute thresholding technique \citep[e.g.][]{Berrilli2005,Schrijver2007} to prepare three different bitmap images: $I_{+}$, $I_{-}$, and $I_{U}$.\\ 
The three bitmap images (see Fig.\Ref{Dscheme}) are:
  \[
    I_{+}(x,y)=\left\{
                \begin{array}{ll}
                  = 1 \quad \textrm{if} \quad M(x,y) \ge 150 \textrm{G}\\
                  = 0 \quad \textrm{\textit{otherwise}}
                \end{array}
              \right.
  \]
  \[
    I_{-}(x,y)=\left\{
                \begin{array}{ll}
                  = 1 \quad \textrm{if} \quad M(x,y) \le 150 \textrm{G}\\
                  = 0 \quad \textrm{\textit{otherwise}}
                \end{array}
              \right.
  \]
  \[
    I_{U}(x,y)=\left\{
                \begin{array}{ll}
                  = 1 \quad \textrm{if} \quad |{M(x,y)}| \ge 150\textrm{G}\\
                  = 0 \quad \textrm{\textit{otherwise}}
                \end{array}
              \right.
  \]

\vspace{\baselineskip}
Eventually, a classical labelling technique identifies and counts the numbers of compact flux density structures (composed of at least two connected pixels) in the three images, $N_{U}$, $N_{+}$, and $N_{-}$, respectively.
The difference $D=(N_{+} + N_{-}) - N_{U}$ counts the number of polarity-inversion line fragments in the magnetogram.

%
%
Figure \ref{fig:R_conf}, right panel, illustrates the correlation between $D$ and four other topological descriptors utilized in this study (namely, wlsg$\_$blos/value$\_$int, mpil$\_$blos/tot$\_$length, mpil$\_$blos/max$\_$length, and mpil$\_$blos/tot$\_$usflux). The panel shows that the correlations are rather weak, thus confirming the different nature of this newly introduced descriptor.

Although the use of an absolute threshold is not always reliable, in our case we have found that the D parameter depends weakly on the variation of a few Gauss around the value adopted. 
When applied on SHARP patches, this technique is extremely fast and provides a new metric that takes into consideration the  complexity of the emerging electrical current system. Moreover, the small dependence on the exact threshold value, due to the topological definition of complexity rather than to an accurate estimate of the magnetic energy reservoir, allows a simpler application in the case of ground-based synoptic telescopes \citep[e.g.][]{Elsworth2015,Gosain2018,Hill2019,Giovannelli2020,Viavattene2020,Calchetti2020}.
\section{Flare forecasting benchmarks}\label{Results}
First, we show the results obtained using separately the R$^*$ and D descriptors, then the results we got using both of them and finally the results compared with other descriptors that can be computed from SDO/HMI magnetograms.

\subsection{Flare forecasting with R$^*$ and D: a heuristic approach}
\begin{figure}[tbp]
\centering
\includegraphics[width=0.48\textwidth]{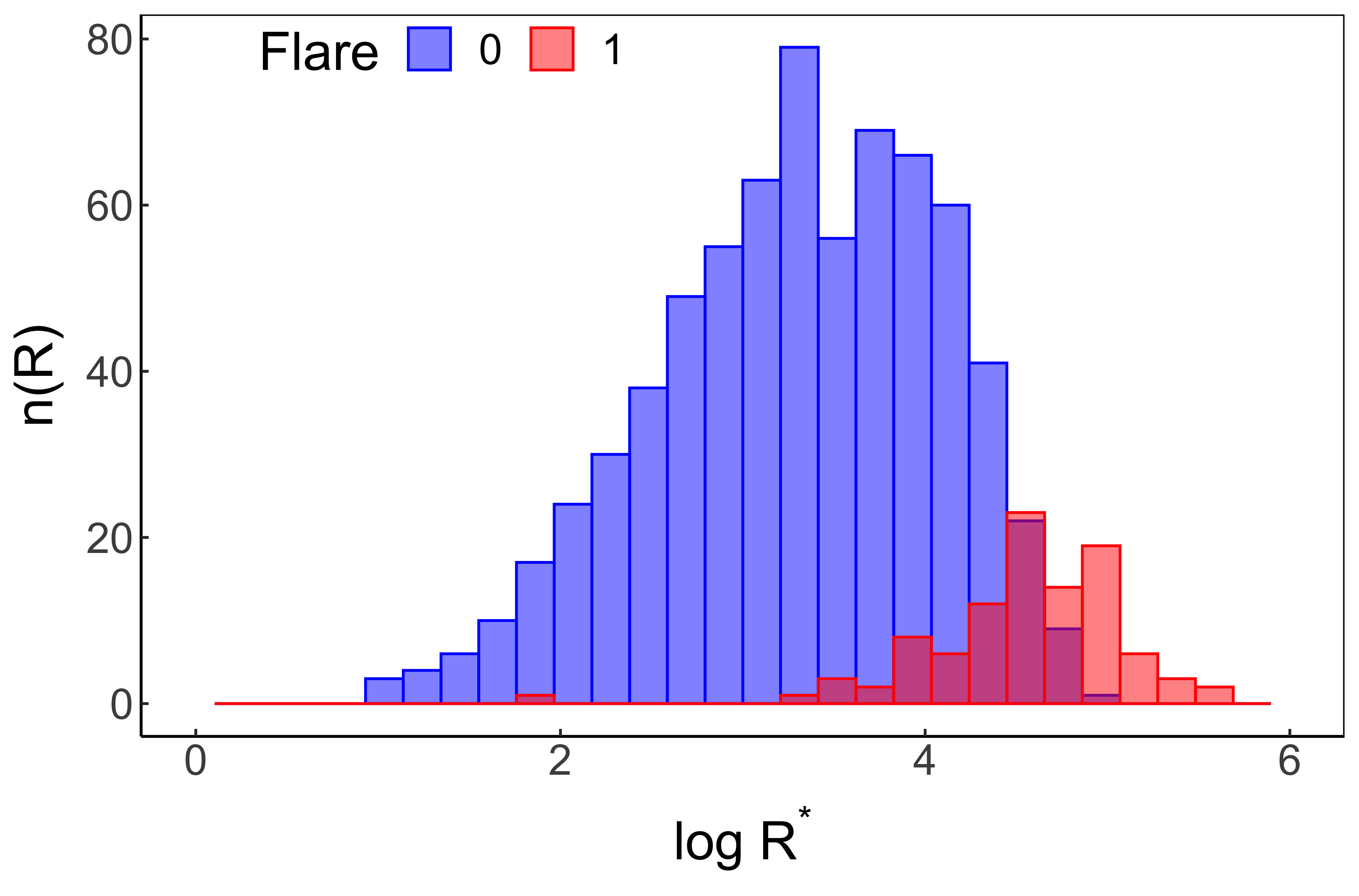}
\includegraphics[width=0.48\textwidth]{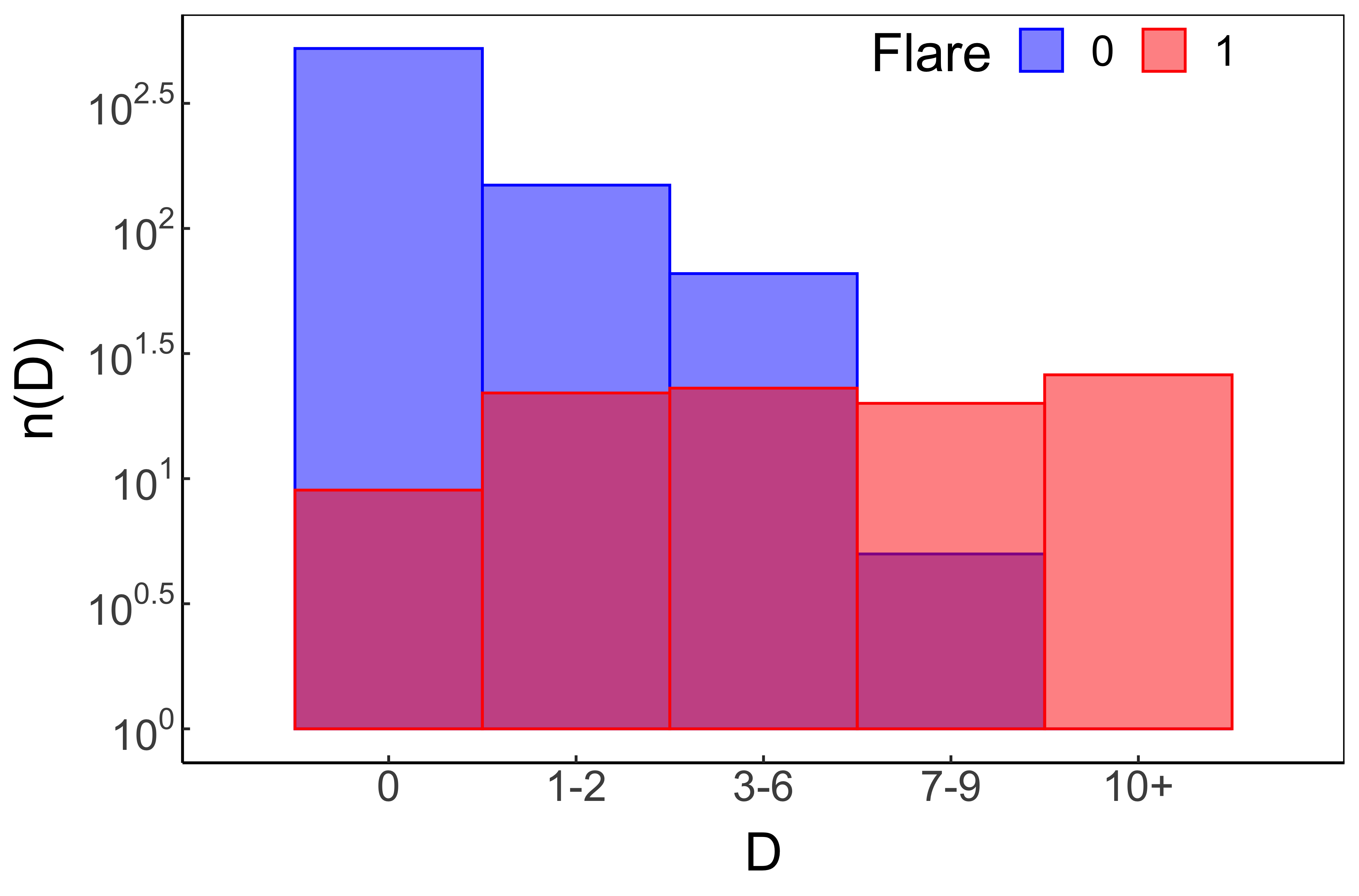}
\caption{Left panel: Histograms of log R$^*$ for all the 845 active regions magnetograms.
The regions with R$^*=0$ are not shown.
Right panel: Histogram of D for the total 845 active regions magnetograms.}
\label{fig:R_D_stats}
\end{figure}

In order to test the effectiveness of R$^*$ and D as statistical classifiers for flare forecasting, we first considered their distributions for all the 845 active regions magnetograms, see Figure \ref{fig:R_D_stats}. In the left panel of Figure \ref{fig:R_D_stats}, we show the distribution of the number of events as a function of log R$^*$.
Almost no M- or X- class flare occurred within 24 hours from the calculation of R$^*$ in a region with log R$^* <3.2$, which includes about
40\% of non-flaring active regions. Only about 
15\% of the regions with log R$^* > 3.2$ show an M- or X- class flare in the next 24 hours.
Among the 100 flaring active regions, 
37\% of them have log R$^* \approx 5$ or greater.

\begin{figure}[htbp]
\centering
\includegraphics[width=0.48
\textwidth]{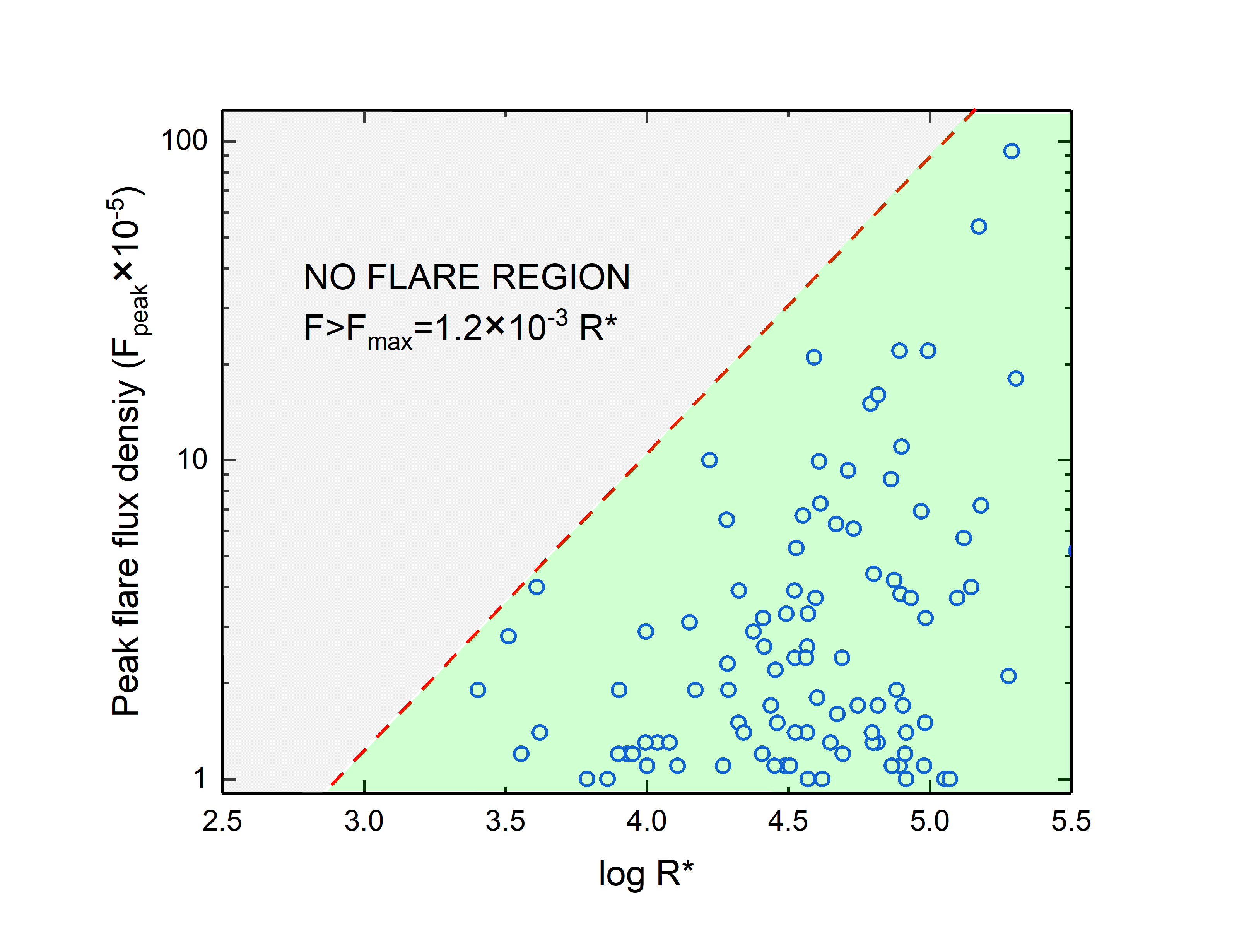}

\caption{
Scatter diagram of the peak flux density in the GOES $0.1-8$ nm band (W m$^{-2}$) vs $\log(R^*)$ for all flaring regions. The red dashed line at $\mathrm{F_{max}=1.2\cdot 10^{-8}R^*}$ defines the green region where all the flares appear. 
We notice that flares from the 24$^{th}$ solar cycle show a similar behavior to those from the 23$^{th}$ one (compare this Figure to Fig.3 in \citealt{Schrijver2007}).}
\label{fig:scatter_R_X_flux}
\end{figure}

The scatter plot in Figure \ref{fig:scatter_R_X_flux} correlates the GOES flux density
with the value of log R$^*$; we notice the same behavior that was found for solar cycle 23 by \cite{Schrijver2007}, with a clustering of flares in the lower right corner.
This result suggests that R$^*$ measures the maximum energy available for flares in a given AR given by $F_{max} = 1.2*10^{-8}$R$^*$. 
As already noted by \citet{Schrijver2007}, it also indicates that ARs reduce their energy through multiple flares, with greater probability of smaller flares: even at very high R$^*$ values, M- class events are far more frequent than X- class.
Comparing the results of this analysis with those shown in \cite{Schrijver2007} is of course an interesting exercise, since the latter ones were obtained from the analysis of ARs appeared during the 23$^{th}$ solar cycle, while those used in the present study came from the 24$^{th}$ one, which was far less active and with a much smaller number of active regions. However, despite these differences between the two cycles, the general trend seems to be very similar (see the exact values in Table \ref{tab:LikeSchrijver}). Namely, there is an almost complete absence of flaring ARs for log R$^* < 3$; about 39\% of the ARs with log R$^* \approx 4.5$ flares in the next 24 hours; the flaring fraction is 93\% for those ARs with log R$^*$ $\approx 5$ or greater.
\begin{table}[hb!]
\begin{center}
\begin{tabular}{lccccc}
\hline
& \multicolumn{5}{c}{log R$^*$} \\
Class           &   log R$^*<$ 3.25  &  3.25 $\leq$ log R$^*$ $<$ 3.75    &   3.75 $\leq$ log R$^*$ $<$ 4.25   & 4.25 $\leq$ log R$^*$ $<$ 4.75    &   log R$^* \geq$ 4.75 \\
 & (\%) & (\%) & (\%) & (\%) & (\%) \\
\hline
$>$M1 & $\sim$ 0 & 3 & 9 & 39 & 93 \\
$>$X1 & 0 & 0 & 0 & 1 & 23 \\
\hline
& \multicolumn{5}{c}{D} \\
Class & D = 0 &  D = 1 \textemdash \, 2 & D = 3 \textemdash \, 6 & D = 7 \textemdash \, 9 & D $\geq$ 10 \\
& (\%) & (\%) & (\%) & (\%) & (\%) \\
\hline
$>$M1 & 2 & 13 & 26 & 80 & 96 \\
$>$X1 & 0 & $\sim$ 0 & 1 & 4 & 26 \\
\hline
\end{tabular}
\caption{Likelihood of M or X flare within 24 hours from the determination of the log R$^*$ and D parameters, similarly to Tab.1 in \cite{Schrijver2007}. 
The bin size for log R$^*$ is 0.5 with the exception of the first and the last bin.
}
\label{tab:LikeSchrijver}
\end{center}
\end{table}

\begin{figure}[htbp]
\centering
\includegraphics[width=0.7\textwidth]{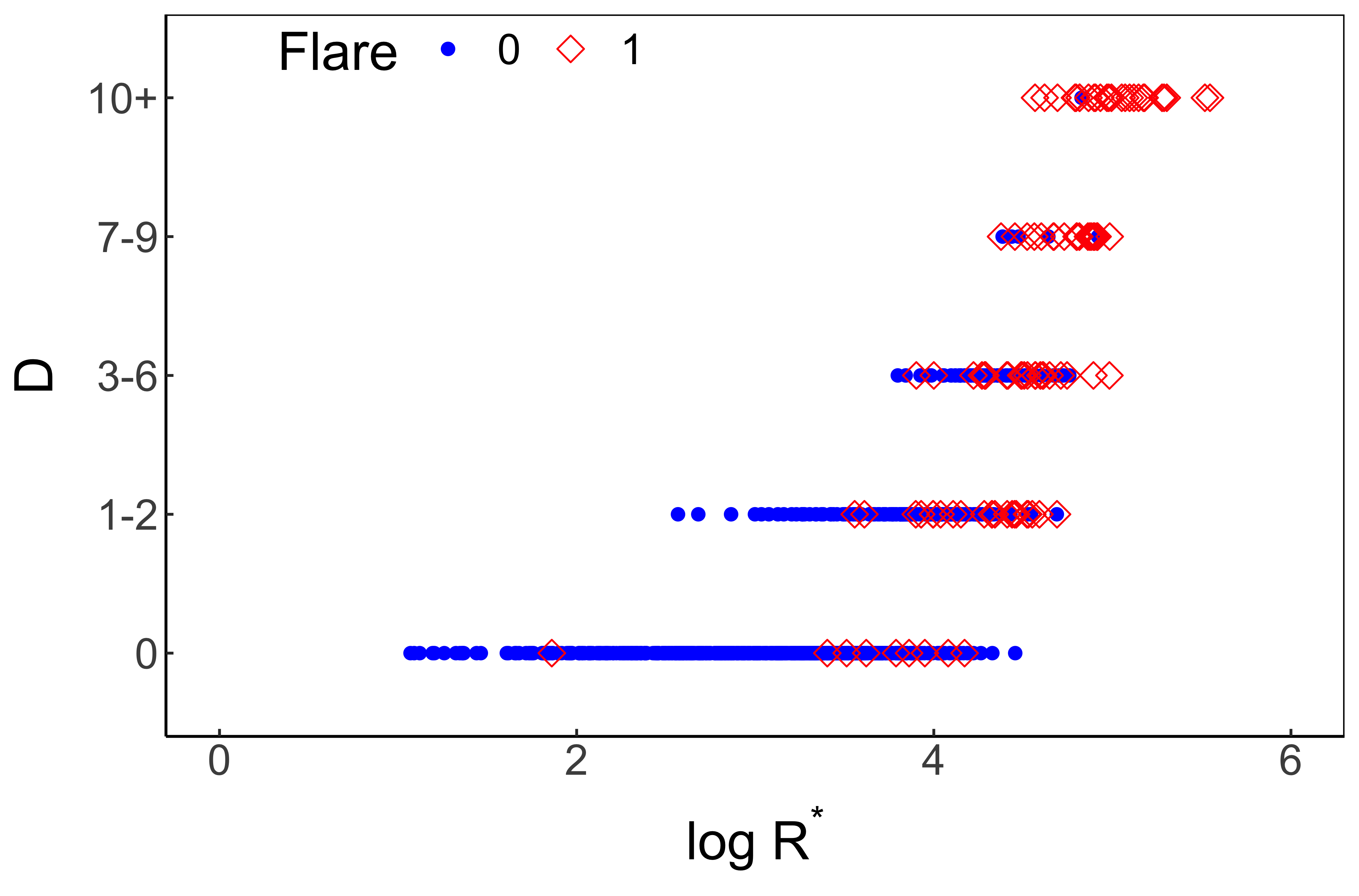}
\caption{Scatter plot of log R$^*$ vs D parameter. 
The flaring ARs are shown with diamonds, while the non-flaring ARs with dots.
ARs with R$^*=0$ are not shown. }
\label{fig:R_vs_D}
\end{figure}

As far as descriptor D is concerned, the right panel of Figure \ref{fig:R_D_stats} shows that about 10\% of the magnetograms in the flaring ARs set (with a flaring probability less than 2\%) and 70\% of the magnetograms in the non-flaring set have D=0. About
41\% of the ARs with D $\geq 1$ have a major flare in the next 24 hours. For D=6, the probability of having a major flare in the next 24 hours grows to 60\%, and for D$\geq 10$ it almost reaches 100\% (see the exact values in Table \ref{tab:LikeSchrijver}).
We completed this preliminary analysis by checking the possible correlation between these two descriptors: Figure \ref{fig:R_vs_D} shows a clear trend of flaring events (represented with red diamond in the plot) for higher values of the two features.

Relying on the results of this preliminary analysis, we investigated the forecasting power of the two descriptors, both separately and jointly.

As a first heuristic approach, we first considered different thresholds of log R$^*$ and D to create deterministic forecasts for M-class flares or more intense in the next 24 hours. We measured the performance of such binary classifiers by means of commonly used classification evaluation metrics.
Specifically, as in \cite{Bobra2015}, we constructed confusion matrices according to the following scheme: we classified as true positives ({\textit{TPs}}) all flaring ARs that have been correctly predicted as flaring; as true negatives ({\textit{TNs}}) all non-flaring ARs that have been correctly predicted as non-flaring; as false negatives ({\textit{FNs}}) all flaring ARs that have been incorrectly predicted as non-flaring; as false positives ({\textit{FP}}s) all non-flaring ARs incorrectly predicted as flaring. 
From these quantities, we could compute the following skill scores to assess the classifying performances of the two descriptors.
The True Positive Rate ({\textit{TPR}}) is defined as:
\begin{equation}
    TPR= \frac{TP}{TP + FN}
\end{equation}
and is the ability of the classifier to identify all of the positive events (in brief, a {\textit{TPR}}$=1$ means that the classifier will recognize all the ARs that had a flare in the next 24h).
We also consider the True Skill Statistics ({\textit{TSS}}):
\begin{equation}
    TSS = \frac{TP}{TP + FN} - \frac{FP}{FP + TN}
\end{equation}
and the Heidke Skill Score ({\textit{HSS}}):
\begin{equation}
    HSS = \frac{2(TP\cdot TN-FN\cdot FP)}{(TP + FN)(TN + FN)+(TP + FP)(TN + FP) =} = 
    \frac{TP+TN-E}{TP+TN+FP+FN-E}
\end{equation}
with 
\begin{equation}
E = \frac{(TP + FN)(TN + FN)+(TP + FP)(TN + FP)}{TP+TN+FP+FN}
\end{equation}
is the expected number of correct random forecasts (either a hit by chance $\frac{(TP + FN)(TN + FN)}{TP+TN+FP+FN}$ or correct rejection by chance $\frac{(TP + FP)(TN + FP)}{TP+TN+FP+FN}$) due
purely to random chance.
These two scores range form -1 to 1 and provide 1 in the case of  perfect forecasts, while for constant forecasts they are equal to 0.
{\textit{TSS}} tends to {\textit{TPR}} when the number of {\textit{TNs}} tends to infinity.
Both {\textit{TSS}} and {\textit{HSS}} can assume negative values if the forecast is worst than a random one.

In Table \ref{tab:performance} we report the values of {\textit{TPR}}, {\textit{TSS}}, and {\textit{HSS}} for the forecast algorithms created using R$^*$ or D alone, and using both R$^*$ and D, varying the threshold settings used to predict whether there will be an M1 or greater flare in the next 24h.
Commenting the results in this table we have the following items.
\begin{table}[tbp]
\begin{center}
\begin{tabular}{l|ccc|ccc|ccc}
\hline
& \multicolumn{3}{c|}{log R$^*$} & \multicolumn{3}{c|}{D} & \multicolumn{3}{c}{log R$^*$ $\&$ D} \\
Threshold           &   $TPR$  &  $TSS$ &  $HSS$ &   $TPR$  &  $TSS$ & $HSS$  &   $TPR$  &  $TSS$ &  $HSS$    \\  
\hline
log R$^*$ $>4$, D $\geq 5$  & 0.85 & 0.66 & 0.42 & 0.58 & 0.55 & 0.62 & 0.58 & 0.55 & 0.62 \\
log R$^*$ $\geq 4.5$, D $\geq 7$  & 0.62 & 0.59 & 0.62 & 0.46 & 0.45 & 0.57 & 0.44 & 0.44 & 0.56 \\
log R$^*$ $\geq 5$, D $\geq 10$  & 0.12 & 0.12 & 0.19 & 0.26 & 0.26 & 0.38 & 0.12 & 0.12 & 0.19 \\
\hline
\end{tabular}
\caption{$TPR$, $TS$ and $HSS$ scores for different M-class flare forecasting algorithms created by setting different R$^*$ and D thresholds.
The three columns in the box identified by log R$^*$ refer to the prediction individually built on $R^*$.
The three columns in the box identified by D refer to the prediction individually built on D.
The three columns in the box identified by log R$^*$ $\&$ D refer to the prediction built with the joint use of log R$^*$ and D.}\label{tab:performance}
\end{center}
\end{table}

\begin{itemize}
    \item Predictor individually built on R$^*$.
When the threshold used to forecast whether an AR will release a flare in the next 24h is log R$^*$ $>4$,  85\% of the positive cases are correctly predicted ({\textit{TPR}}), however, {\textit{TSS}} and {\textit{HSS}} present smaller values.
The performance improves with higher threshold and, for log R$^*$ $\geq4.5$, it is optimal for an M class flare predictor.
    \item Predictor individually built on D alone. 
Again, for a threshold set at D $\geq 5$, the predictor guesses $58\%$ of the flaring ARs, with a relatively small number of false positive ({\textit{TSS}}$=0.55$).
With a threshold of D $\geq 7$, the skill scores decrease slightly; even more increasing the D threshold, the classifier correspondingly increases its effectiveness in selecting only flaring ARs, but it recognizes only $26\%$ of all the flaring ARs.
Thus, its overall performances decrease to $TSS=0.26$ and $HSS=0.38$.
    \item Predictor built on both R$^*$ and D.
For thresholds set at log R$^*$ $\geq4$ and D $\geq 5$, $58\%$ of the flaring ARs are correctly recognized.
Apparently, the algorithm is not as successful in retrieving all the flaring ARs, since its {\textit{TPR}} values are smaller than or at most equal to the {\textit{TPR}} values of the classifiers built on R$^*$ or D alone, for the same threshold values.
Nevertheless, the high values of the {\textit{TSS}} and {\textit{HSS}} scores demonstrate that this predictor performs highly satisfactorily, particularly for the thresholds log R$^*$ $\geq4$ and D $\geq 5$.
\end{itemize}

\subsection{Flare forecasting with R$^*$ and D: an ML approach}
The previous approach to the computation of the prediction effectiveness has the advantage to assess the role of the two predictors while tuning the threshold.
However, the way these threshold values are chosen is arbitrary and therefore it is necessary to perform an optimized calculation of parameters by means of a machine learning approach that allows an automatic computation of the scores.
Therefore, we have studied the performances of a hybrid LASSO supervised algorithm \citep{Benvenuto2018} in which the classification is obtained by means of the application of a fuzzy clustering technique on the outcomes of the LASSO regression step. More specifically, the fuzzy clustering algorithm automatically computes the threshold separating the regression outcomes into two classes, corresponding to flare/no-flare prediction. The regression method is therefore transformed into a classifier whose binary outcomes are used to compute the contingency tables.
In order to realize the training phase we have considered the same data sets of ARs used in \cite{Campi2019}. Specifically:
\begin{itemize}
    \item We have exploited the SDO/HMI archive in the time range between September, 14 2012 and April, 30 2016.
      \item The corresponding magnetograms have been grouped into four subsets belonging to the four issuing times 00:00, 06:00, 12:00, 18:00 UT. Of course, it may happen that an AR lasted more than one day: in that case, each one of the four subsets will contain as many samples associated to that AR, as the days the AR lasts.
    \item For each subset, i.e., for each issuing time, we randomly extracted two thirds of ARs in order to construct the training set using the samples contained in that subset while the remaining one third was used to populate the corresponding test set.
\item Each sample in each training set has been labelled by using GOES data in such a way that label "1" corresponds to an event occurrence within the 24 hours from the issuing time.
    
\end{itemize}
This process was repeated $100$ times, so that, at the hand, we had $100$ realizations of the training set and $100$ realization of the test set, for each issuing time.
From each AR we extracted D and log R$^*$ and fed the ML algorithm with the corresponding two-dimensional feature vectors.
In order to assess the forecasting performances we computed the same skill score as above and the results we obtained are in Table \ref{tab:ranking_tss_soloR_D}.
Differently than what done for Table \ref{tab:performance}, Table \ref{tab:ranking_tss_soloR_D} is populated by means of a neural network which has been applied against a test set with no intersection with the training set. 
Further, the training set used to optimize hybrid LASSO is significantly more imbalanced against M flares with respect to the data set utilized to generate Table \ref{tab:performance}.
However, the score values automatically obtained in Table \ref{tab:ranking_tss_soloR_D} are in several cases comparable with the ones obtained by heuristically tuning the thresholds on log R$^*$ and D. 

\begin{table}[tbp]
\begin{center}
\begin{tabular}{l|ccc}
\hline
& \multicolumn{3}{c}{Test set  - at least M1 flares}\\
 \hline
& \multicolumn{3}{c}{log R$^*$ $\&$ D} \\
Issuing time         &   $TPR$  &  $TSS$ &  $HSS$    \\  
\hline
00:00:00UT  & 0.47 $\pm$ 0.13 & 0.40 $\pm$ 0.12 & 0.22 $\pm$ 0.06\\
06:00:00UT  & 0.54 $\pm$ 0.09 & 0.47 $\pm$ 0.08 & 0.31 $\pm$ 0.05\\
12:00:00UT &  0.49 $\pm$ 0.09 & 0.44 $\pm$ 0.08 & 0.33 $\pm$ 0.05\\
18:00:00UT &  0.43 $\pm$ 0.04 & 0.38 $\pm$ 0.04 & 0.30 $\pm$ 0.03\\
\hline
\end{tabular}
\caption{Average $TPR$, $TSS$ and $HSS$ values and corresponding standard deviations for the outcomes of hybrid Lasso when applied to 100 random realizations of the
training/test sets at four specific forecast issuing times.} \label{tab:ranking_tss_soloR_D}
\end{center}
\end{table}

We quantify the impact on the forecasting performance of the two descriptors introduced in this study when used in combination with many more predictors. The performed feature ranking analysis relies again on hybrid LASSO, and follows the approach described by \cite{Campi2019}.
That paper used as properties the $171$ ones computed within the Horizon 2020 FLARECAST project (\url{http://flarecast.eu}) and arranged them again in the four training sets corresponding to the four issuing times (00:00, 06:00, 12:00, 18:00 UT); then, we applied a hybrid LASSO machine supervised machine learning technique \citep{Benvenuto2018} in order to both realize binary prediction within the next 24 hours for each issuing time, and identify the features that mostly impact such prediction. In the present paper we applied the same algorithm to the same data set, this time enriched with descriptors R$^*$ and D. Coherently with the previous sections, the task was the prediction of M1 or more intense events within the next 24 hours. 

\begin{table}[tbp]
\begin{center}
\begin{tabular}{l|ccc|ccc}
\hline
& \multicolumn{6}{c}{Test set  - at least M1 flares}\\
 \hline
& \multicolumn{3}{c|}{171 descriptors}
& \multicolumn{3}{c}{171  descriptors + log R$^*$ $\&$ D} \\
Issuing time         &   $TPR$  &  $TSS$ &  $HSS$   &   $TPR$  &  $TSS$ &  $HSS$ \\  
\hline
00:00:00UT  & 0.66 $\pm$ 0.16 & 0.56 $\pm$ 0.14 & 0.27 $\pm$ 0.06 & 0.45 $\pm$ 0.21 & 0.57 $\pm$ 0.14 & 0.34 $\pm$ 0.10 \\
06:00:00UT  & 0.76 $\pm$ 0.06 & 0.67 $\pm$ 0.05 & 0.35 $\pm$ 0.04 & 0.74 $\pm$ 0.07 & 0.65 $\pm$ 0.06 & 0.34 $\pm$ 0.04 \\
12:00:00UT & 0.74 $\pm$ 0.07 & 0.66 $\pm$ 0.06 & 0.38 $\pm$ 0.04 & 0.73 $\pm$ 0.07 & 0.65 $\pm$ 0.06 & 0.38 $\pm$ 0.04\\
18:00:00UT & 0.71 $\pm$ 0.08  & 0.64 $\pm$ 0.07 & 0.39 $\pm$ 0.04 & 0.70 $\pm$ 0.08 & 0.63 $\pm$ 0.07 & 0.38 $\pm$ 0.04 \\
\hline
\end{tabular}
\caption{Average $TPR$, $TSS$ and $HSS$ values and corresponding standard deviations for the outcomes of hybrid Lasso when applied to 100 random realizations of the
training/test sets at four specific forecast issuing times if the 171 predictors of \cite{Campi2019} (left panel) and the 171 predictors plus log R$^*$ and D (right panel) are used.} \label{tab:skill_score_con_flarecast}
\end{center}
\end{table}

\begin{figure}[tbp]
\centering
\includegraphics[width=0.8\textwidth]{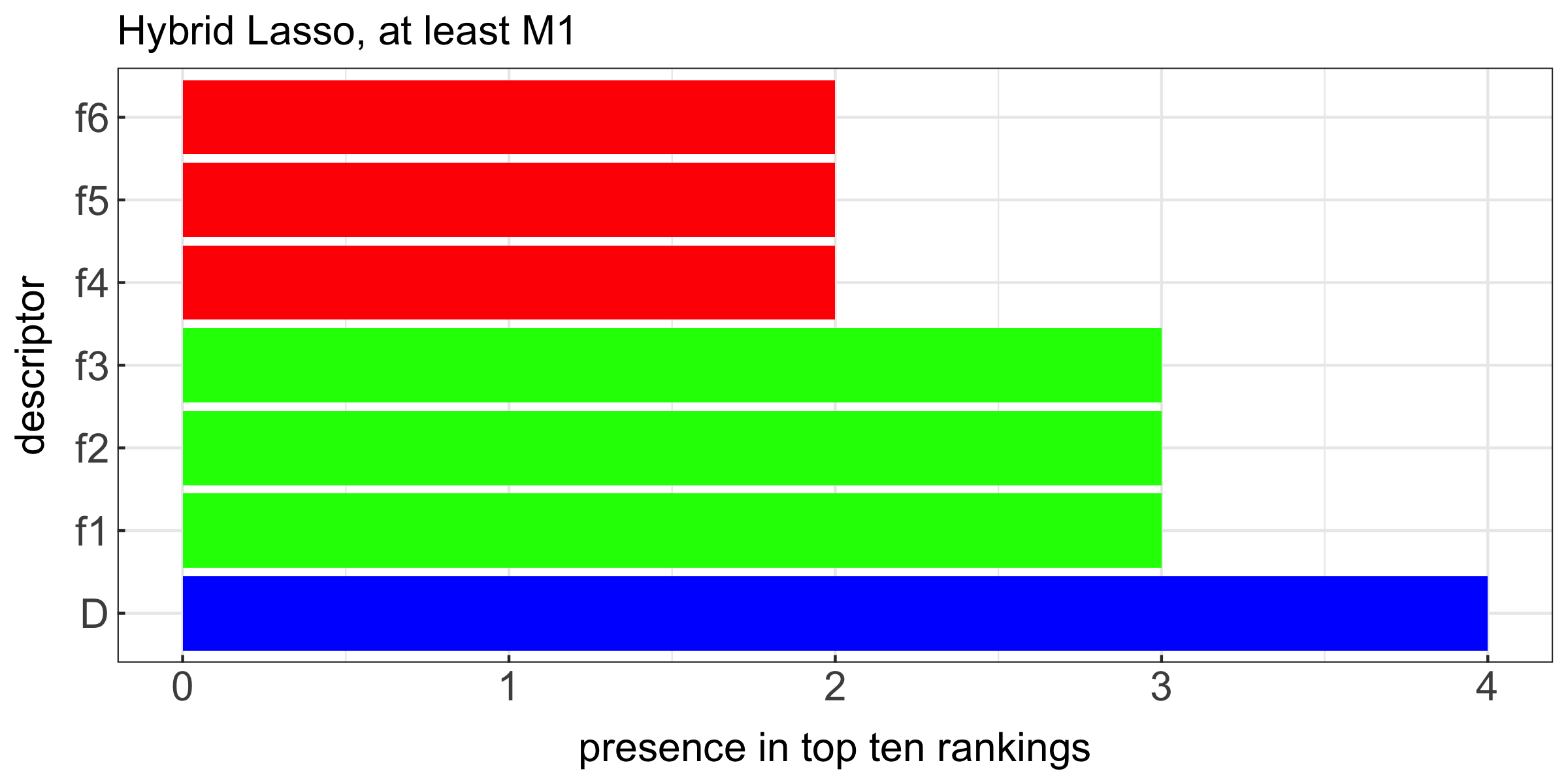}
\caption{Histograms counting the number of times predictors  are in the top-10 rankings, on average over the 100 random realizations of the training and test sets. Descriptors from ``f1'' to ``f6'' are  ``sfunction\_blos/zq'', ``sharp\_kw/snetjzpp/total'', ``sharp\_kw/ushz/stddev'',``decay\_index\_br/maxl\_over\_hmin'', ``helicity\_energy\_bvec/abs\_tot\_dedt\_in'',``wlsg\_br/value\_int'', respectively, and their description can be found in Table \ref{tab:descriptors_definition}.                  }\label{fig:features_ranking}
\end{figure}

\begin{figure}[tbp]
\centering
\includegraphics[width=0.8\textwidth]{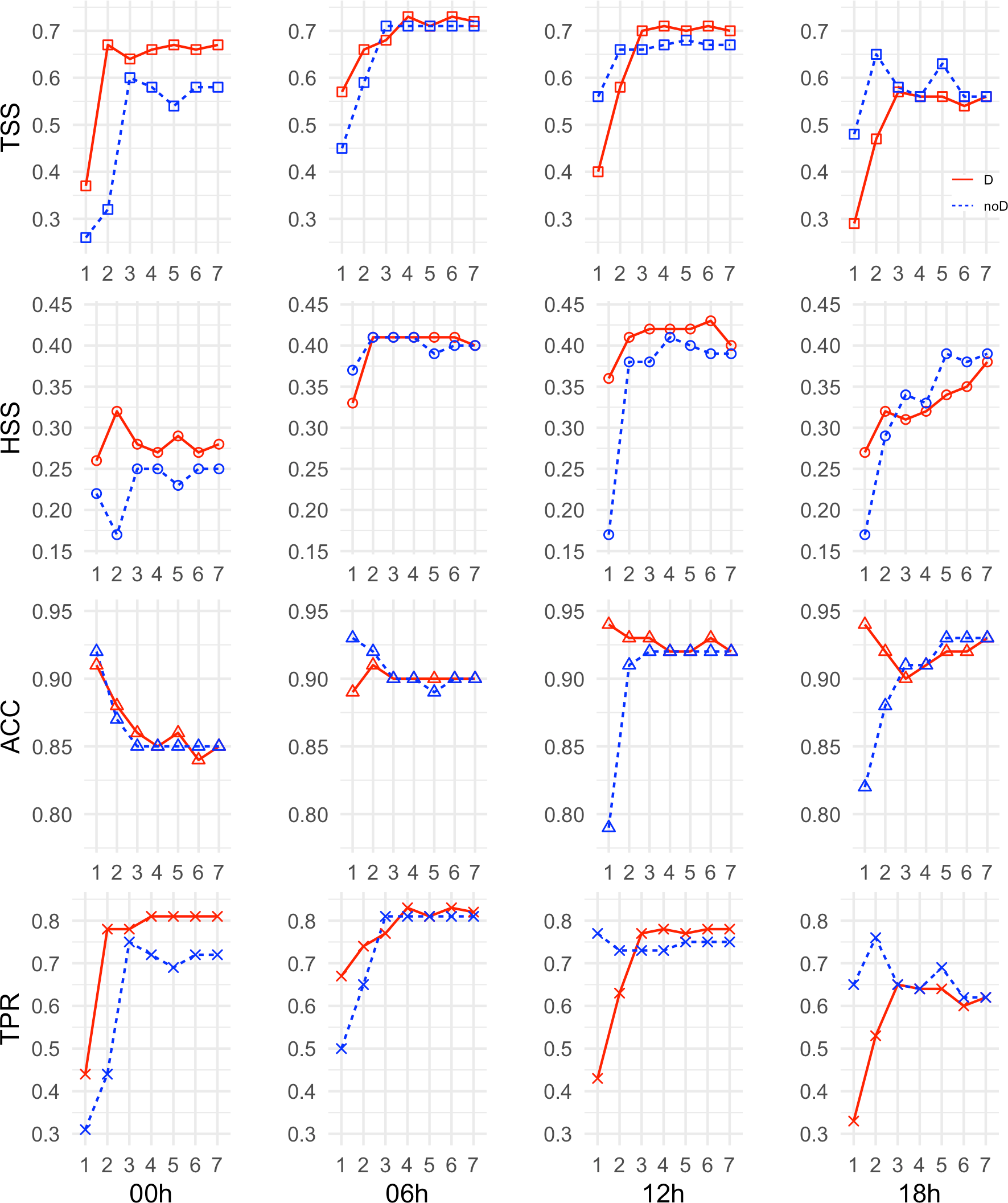}
\caption{Skill scores obtained by using just the 7 features with the best rankings in decreasing order, from 1 to 7, for both machine-learning methods and all four issuing times. Features are added one at a time. The red solid line refers to the case where $D$ is added as first descriptor, the blue dashed line refers to the case without $D$.}\label{fig:featureXfeauture}
\end{figure}

The results of binary classification are shown in Table \ref{tab:skill_score_con_flarecast}. Comparing Table \ref{tab:ranking_tss_soloR_D} and \ref{tab:skill_score_con_flarecast}, one can easily comment that, on the one hand, the use of more descriptors allows machine learning to produce higher scores. On the other hand, comparing the two blocks of columns within Table \ref{tab:skill_score_con_flarecast} suggests that the use of the two new descriptors does not significantly increase the score values.
However, the results of the feature ranking analysis illustrated in Figure \ref{fig:features_ranking} shows that the descriptor $D$ plays a significant role in the realization of the prediction performances.
In particular, the histograms in the figure describe the predictors in the training set to which LASSO assigns the highest weighs for the prediction task.
In this ranking, descriptor $D$ is the only one among the $173$ overall predictor, which is always present for all four issuing times. In order to quantitatively assess the contribution of $D$ to a multi-predictor forecast, we trained hybrid Lasso on $100$ random realizations of the training and test sets according to the following process: first, the training step relies just on $D$ and the we added the other features one by one, according to the order given by the ranking. The resulting skill scores are then compared to the ones obtained by training the network with training sets that do not contain D and in which the features are added one by one, again according to the ranking. The scores averaged over the $100$ random realizations, are shown with in the panels of Figure \ref{fig:featureXfeauture}, with the solid red line when $D$ is included in the training set as first descriptor, and with the dashed blue line when $D$ is not in the training set. In this figure we also considered the values of the accuracy {\textit{ACC}}, which is defined as
\begin{equation}\label{accuracy}
ACC = \frac{TP + TN}{TP + TN + FP + FN}.
\end{equation}

We notice that, for most cases, when D is used for the training, either the outcomes lead to higher skill score values, or the saturation of the skill score profiles is reached in advance.

As far as descriptor $R^*$ is concerned, we found that its position in the averaged feature ranking ranges between rank, $20$ and rank $30$ for the four issuing times, while in the analysis in \cite{Campi2019} the two Schrijver's R properties (one for the B$_{los}$ and one for B$_{radial}$ magnetograms, \citealt{Guerra2018}) were ranked between the 80th and 110th position among the $171$ features considered in that study. 
We interpret this relevant ascent in the ranking as the result of the larger information content of R$^*$ compared to that in R, obtained by exploiting the full resolution of HMI magnetograms.

\begin{table}[tbp]
\begin{center}
\begin{tabular}{|c|c|}
\hline
Descriptor & Definition \\
\hline
``wlsg\_blos/value\_int'' and  ``wlsg\_br/value\_int'' & B$_{los}$ and B$_{radial}$ Falconer’s $^L$WL$_{SG}$ \\ 
``mpil\_blos/tot\_length'' & Total length of all MPILs\\ ``mpil\_blos/max\_length'' & Maximum length of a single MPIL\\ 
``mpil\_blos/tot\_usflux'' & Total unsigned flux around all MPILs\\
``sfunction\_blos/zq'' & tifractal structure function inertial range index\\
``sharp\_kw/snetjzpp/total''&Sum of absolute value of  net currents per polarity\\ ``sharp\_kw/ushz/stddev''& Standard deviation of the unsigned vertical current helicity\\
``decay\_index\_br/maxl\_over\_hmin''& Maximum ratio of MPIL length to minimum height of critical decay index\\ 
``helicity\_energy\_bvec/abs\_tot\_dedt\_in''& Absolute value net vertical Poynting flux\\
\hline
\end{tabular}
\caption{List and definitions of the mentioned descriptors from the FLARECAST project.}\label{tab:descriptors_definition}
\end{center}
\end{table}

\section{Conclusions}
This study introduced two novelties concerning the computation and exploitation of AR descriptors for use with prediction purposes. First, we developed a new release of the algorithm for the computation of the Schrijver R value. In fact, among AR descriptors, this value plays a well-established role, this parameter being a measure of the magnetic flux associated to PILs in the solar photosphere. The original algorithm for the computation of R was designed for MDI magnetograms and unreliably adapts to data gathered by the new generation HMI cameras. In this paper, we modified this algorithm in order to account for the differences in the design of the two instruments and therefore to perform an accurate computation of the descriptor. To test whether the rescaled algorithm, designed to work on the full resolution magnetograms acquired by SDO/HMI, estimated the flare occurrence probabilities in a comparable way with \cite{Schrijver2007}, we selected 845 magnetograms (100 of flaring regions and 745 of non-flaring regions) and repeated the same statistical analysis. We found that the occurrence rates of M- or X-class flares for the new R$^*$ value is consistent with what is reported in \cite{Schrijver2007}.

The second contribution of the paper relied on the well-established observation that topologically complex ARs are strongly correlated to flaring emissions. Therefore, we proposed a topological descriptor that counts the number of separated PILs fragments in the ARs and whose computation relies on a fast thresholding/labelling process.

The rest of the study was devoted to bench-marking these two descriptors and, more specifically, to assess their impact on the flare prediction process by using both a heuristic thresholding procedure and an ML approach. We considered SDO/HMI magnetograms in the time range between September 14 2012 and April 30 2016 and grouped them into four subsets belonging to the four time stamps 00:00, 06:00, 12:00, 18:00 UT. For each subset we created a set made of two thirds of randomly extracted ARs that represented the training set for the supervised ML method we utilized for the analysis. In particular, the use of a sparsity-enhancing technique for classification allowed the identification of the ARs descriptors that mostly impact the flare forecasting process. This analysis showed that $D$ is the only one among almost two hundred predictors that achieves the top-ten ranking for all four issuing times considered, and that its use in the training set implies either a steeper increase of the skill score profiles or the achievement of higher skill scores values. This result is a quantitative confirmation of phenomenological analyses of the correlation between AR complexity and the probability of flare emission and, also, a nice validation of the reliability of ML-based feature ranking procedures.

In the case of $R^*$, the outcome of the feature ranking process showed that this new way to compute this descriptor analogously increases its position in the ranking with respect to the same descriptor computed according to the approach proposed in \citep{Schrijver2007}. However, this improvement is not as significant as for $D$ and this is the reason why the use of $R^*$ does not lead to higher skill score values.

Descriptors R$^*$ and D are in the process of being implemented in an updated version of the Space WEatheR TOr vergata university (SWERTO) service \citep{Berrilli2018IAU, Napoletano2018, DelMoro2018annals, SWERTO2019} along with other Space Weather event predictors.

\section*{Acknowledgement}
\label{sec:acknowledge}
This paper is based on Master's thesis research "\textit{The calibration of statistical solar flare  methods for images from SDO/HMI space instrument}" conducted by D. Cicogna under the supervision of Prof. F. Berrilli, submitted to Physics Graduate Program at the University "Sapienza" of Rome, May 2018. D. Cicogna thanks Prof. Silvia Masi for the guidance during the thesis work. SWERTO space weather service has been financed by the Regione Lazio FILAS-RU-2014-1028 grant (November 2015 – October 2017). This research is partially supported by the Italian MIUR-PRIN grant 2017APKP7T on \textit{Circumterrestrial Environment: Impact of Sun-Earth Interaction}. D. Calchetti was supported by the Joint Research PhD Program in “Astronomy, Astrophysics and Space Science” between the universities of Roma Tor Vergata, Roma Sapienza and INAF. F. Benvenuto, C. Campi, S. Guastavino and M. Piana acknowledge the financial contribution from the agreement ASI-INAF n.2018-16-HH.0. The authors are grateful to the anonymous referee for the careful checking of the manuscript and for helpful comments that improved the paper.

\bibliography{cicogna}
\bibliographystyle{aasjournal}

\end{document}